\documentclass[twocolumn, fp]{jpsj3}
\usepackage{amsmath}
\usepackage{amssymb}
\usepackage{txfonts}
\usepackage{bm}
\usepackage{subfigure}

\title{Electrodynamics in Skyrmions Merging}

\author{Rina TAKASHIMA\thanks{takashima@scphys.kyoto-u.ac.jp} and Satoshi FUJIMOTO}
\inst{Department of Physics, Kyoto University, Kyoto 606-8502, Japan}

\abst{
In a recent study the coalescence of magnetic skyrmions was observed in a metallic chiral magnet Fe$_{0.5}$Co$_{0.5}$Si when the skyrmion phase is destroyed, and numerical simulations demonstrated the existence of a monopole at the merging point of two skyrmion lines. The exchange interaction between such magnetic textures and the conduction electrons can be described by emergent electromagnetism. In this paper, we investigate the effect of a skyrmions-merging process on conduction electrons by calculating induced electric currents.  
Here, in addition to the exchange interactions, we consider the antisymmetric spin-orbit couplings (SOC) due to broken inversion symmetry, which is an essential ingredient for the realization of skyrmion texture in the itinerant magnet Fe$_{0.5}$Co$_{0.5}$Si. We obtain an adiabatic current which is dissipationless, and dissipative currents driven by the effective electromagnetic fields including the effect of SOC. In terms of the effective fields, a moving monopole at the merging point turns out to be a dyon-like object; i.e. it has both electric charge and magnetic charge. 
 }

\begin{document}

\maketitle

\section{Introduction}
\begin{figure}[t]
	\begin{minipage}{0.5\columnwidth}
		\begin{center}
			\includegraphics[ clip, width=0.9\columnwidth]{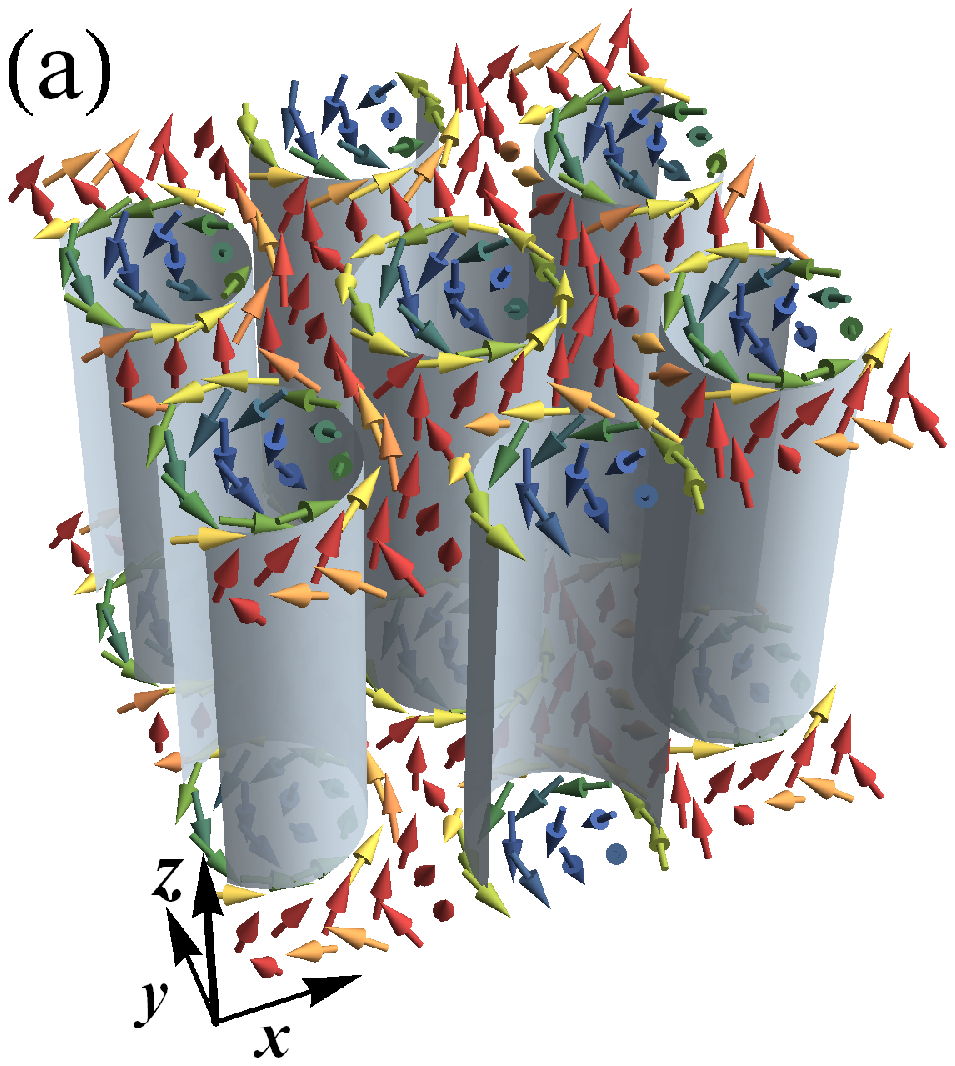}
		\end{center}
	\end{minipage}\begin{minipage}{0.5\columnwidth}
		\begin{center}
			\includegraphics[clip, width=0.8\columnwidth]{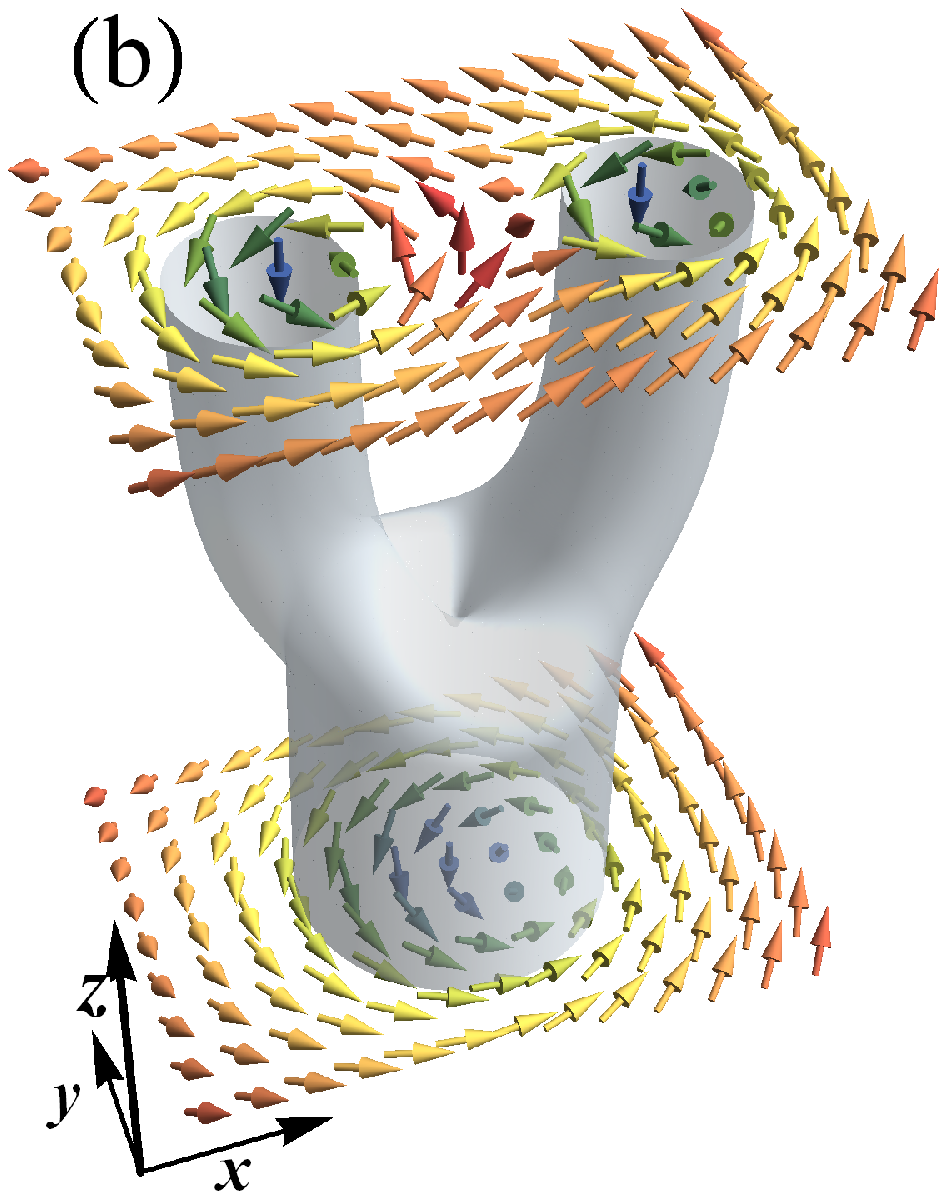}
		\end{center}
	\end{minipage}
		\caption{(Color online) The illustration of three-dimensional spin fields. The arrows indicate the directions of the spins, and {the curved surface corresponds to the area where the spins are parallel to the $xy$ plane; i.e. $n_z =0$. This surface characterizes the profile of skyrmions.} (a) The schematic illustration of a skyrmion lattice in a three-dimensional system. It is translationally invariant in the $z$ direction.
(b) A snapshot of the dynamics of the exchange field in our model. Two skyrmion lines coalesce into one skyrmion line with a monopole at the merging point. {The monopole moves along the positive $z$ direction, leading to the coalescence of skyrmions.} 
}\label{fig:up}
\end{figure}

Dynamics of topologically stable objects such as vortices often induce novel properties in a wide range of phenomena.
In a certain class of superconductors, for example, dynamics of vortices changes the electromagnetic properties.
Such topologically stable objects are also realized in chiral magnets as magnetic skyrmions. Magnetic skyrmions have been intensively studied both experimentally\cite{Muhlbauer2009, Yu2010, Munzer2010, Yu2011, Nagaosa2013} and theoretically\cite{Zang2011, Lin2013a, Iwasaki2013}. The formation of a lattice of skyrmion lines was observed in three-dimensional materials such as MnSi \cite{Muhlbauer2009} and Fe$_{1-x}$Co$_{x}$Si. \cite{Munzer2010} (See Fig. \ref{fig:up} (a) for an illustration).

Recently a new kind of dynamics was observed when the skyrmion phase is destroyed 
in a metallic chiral magnet Fe$_{0.5}$Co$_{0.5}$Si.\cite{Milde2013} 
On the surface of the sample, it was observed that two skyrmions coalesce into one elongated skyrmion, which implies the existence of a defect of the magnetic texture because of a nontrivial topology of a skyrmion; 
a single skyrmion cannot be removed by continuous deformation without any defects.  
The numerical simulation clarified the dynamics of spins in a bulk crystal, and found the creation and motion of a point defect with a non-trivial texture, a hedgehog monopole, at the merging point. This dynamics of skyrmions and monopoles is supposed to lead new phenomena.

Skyrmions have attracted attention not only for the topological stability but also for their interactions with conduction electrons.
In the presence of strong exchange interaction, topological magnetic textures, such as skyrmions and monopoles, can be sources of emergent electromagnetic fields, giving rise to nontrivial Berry curvatures\cite{Volovik1987, Xiao2010}.
Such emergent electromagnetic fields act on conduction electrons just like the electromagnetic field. 
The topological Hall effect, for example, is induced by the emergent magnetic field of skyrmions\cite{Bruno2004}, since a skyrmion carries one quantum of emergent magnetic flux. Furthermore a hedgehog monopole is regarded as a magnetic monopole, i.e., it carries one quantum of emergent magnetic flux.\cite{Milde2013}  

Such emergent electromagnetism in the presence of above dynamics is expected to be further interesting, especially when we take into account the antisymmetric spin-orbit coupling (SOC), which leads to the Dzyaloshinskii-Moriya (DM) interaction {\cite{Dzyaloshinsky1958, Moriya1960}}. 
The DM interaction induces the formation of skyrmions in chiral magnets such as Fe$_{1-x}$Co$_{x}$Si and MnSi. 
A recent study showed that DM interactions originate from the combined effects of the exchange interaction and the antisymmetric SOC due to broken inversion symmetry of the crystal structure. \cite{Freimuth2013}
In general, it is also known that an antisymmetric SOC causes intriguing properties such as an adiabatic current and an anomalous velocity \cite{Nagaosa2010}.   
Since Fe$_{1-x}$Co$_{x}$Si and MnSi are itinerant magnets, the effect of the antisymmetric SOC on conduction electrons needs further investigations.

In this paper, we investigate electromagnetic effects on conduction electrons induced by skyrmions-merging dynamics with a monopole as shown in Fig. \ref{fig:up} (b), where the exchange interactions and antisymmetric SOC are considered.  
Especially we calculate the electric current and the effective electromagnetic fields induced by the above dynamics assumed to be an adiabatic process.
We demonstrate that the above process drives an adiabatic current which is dissipationless due to the antisymmetric SOC. Furthermore  in terms of the effective field we obtained, the moving hedgehog monopole turn out to be a dyon\cite{Rajaraman1983}-like object; i.e. it has both electric and magnetic charges. This remarkable property is due to the dynamical effect; the antisymmetric SOC gives rise to the magnetoelectric effect. 

We note that the effective fields due to antisymmetric SOC, especially Rashba SOC, under the strong exchange interaction has been studied by several authors.\cite{Kim2012a,Tatara2013,Nakabayashi2014} We emphasize that the main purpose of this paper is to point out the above-mentioned novel effect due to the interplay between the dynamics of topological objects and the magnetoelectric effect raised by the antisymmetric SOC, which has not been reported so far.

The remainder of this paper is organized as follows.
In Sect. 2 we will describe the system and derive a dissipationless current, which are independent of the relaxation time, by using the semiclassical theory of a wave packet.
In  Sect. 3 we calculate the effective electromagnetic field and a dissipative currents induced by the effective electromagnetic field within the quasiclassical approximation.
In Sect. 4 the obtained effective fields and currents are evaluated for the process where two skyrmions merge with a monopole. It is found that the monopole behaves like a dyon in this process. 
Finally, we summarize the results in Sect. 5.
\section{Dissipationless current}
We consider conduction electrons in a three-dimensional chiral magnet described by the Hamiltonian
\begin{align}
H&=\frac{\hat{\bm p}^2}{2m}
-J\bm{M}(\hat{\bm r} ,t)\cdot\bm{\sigma}
+\alpha_{so}\bm{g}\left(\hat{\bm p}\right)\cdot\bm{\sigma} 
+v_{\rm{imp}}(\hat{\bm{r}}),
\label{eq:H}
\end{align}
where $\bm{\sigma}$ is the vector of Pauli matrices in the spin space. 
The second term is the exchange interactions between the conduction electrons and the exchange field $\bm{M}(\hat{\bm r} ,t)$ due to the magnetization, with the coupling constant $J>0$. We assume that the magnitude of the exchange field is constant and only the direction varies slowly in space and time;  $\bm{M}(\hat{\bm r} ,t)=M\bm{n}(\hat{\bm r} ,t)$ with $ \bm{n}(\hat{\bm r} ,t)=(\sin\theta\cos\phi, \sin\theta\sin\phi, \cos\theta)$.
The third term is the antisymmetric SOC, with $\alpha_{so}$ being the coupling constant, and $\bm{g}_{so}(\hat{\bm{p}})$ is the SOC field which arises due to the crystal structure without inversion symmetry. In the last term, we include an impurity potential $v_{\rm{imp}}(\bm r)$, which is assumed to be isotropic and independent of spins.

In general, $\bm{g}_{so}(\hat{\bm{p}})$ is given by the average of the operator $\hat{\bm{p}}\times \nabla V(\hat{\bm r})$, where $V(\hat{\bm r})$ is a potential due to broken inversion symmetry, which depends on the detail of the crystal structure.\cite {Fujimoto2012} We, here, determine the form of $\bm{g}_{so}(\hat{\bm{p}})$ from the symmetry argument as follows.\cite{Fujimoto2012, Frigeri2004}
Firstly,  $\bm{g}\left(\hat{\bm p}\right)\cdot\bm{\sigma}$ is required to be even under a time reversal transformation and to be odd under a parity transformation; therefore we obtain the condition, $\bm{g}(-\hat{\bm{p}})=-\bm{g}(\hat{\bm{p}})$, by noting that the spin is odd under a time reversal transformation and even under a parity transformation.
Secondly, we assume that $\bm{g}\left(\hat{\bm p}\right)\cdot\bm{\sigma}$ is invariant under all the symmetry operations of the point group which the crystal belongs to. MnSi, Fe$_{0.5}$Co$_{0.5}$Si and FeGe, where skyrmions are experimentally observed, have the cubic space group P2$_{1}$3. The corresponding point group is the tetrahedral group $T$, which have neither inversion symmetry nor mirror symmetry.
Assuming that the electron density is low, we just take the lowest order in $\bm p$, $\bm{g}(\hat{\bm{p}})=  (\hat p_x, \hat p_y, \hat p_z)$. 
This simple form can give rise to DM interactions which induce a helical ordered state and a skyrmion lattice state.

In this section we will calculate dissipationless currents driven by skyrmions-merging dynamics. Since the Hamiltonian depends on the exchange field $\bm n(\hat{\bm r} ,t)$, which slowly varies in space and time, we adopt the semiclassical theory with wave packet dynamics.\cite{Xiao2009}
Wave packet dynamics describes well such a situation by including Berry curvatures.  
Recently Freimuth \textit{et al.}\cite{Freimuth2013} applied this formalism to chiral magnets, and derived a DM interaction and an electric charge of skyrmions in a static skyrmion lattice.  
Here we calculate electronic current density in a dynamical situation.

We note that the DM interaction, which appears in the free energy density of the magnetization, is proportional to the coupling constant $\alpha_{so}$ in Eq. ({\ref{eq:H}}), and it can be confirmed by calculating the free energy density semiclassically using the Berry curvatures.{\cite{Freimuth2013}}   
Since the wavelength of a helix or the size of a skyrmion is linear in the inverse of the DM coefficient, 
$\alpha_{so}^{-1}$ can be assumed to be the first order of the characteristic length of the magnetic texture. Thus, we introduce the characteristic length scale $l_{sk}$, which is written as 
\begin{equation}
\alpha_{so} = \left(\frac{1}{k_F l_{sk}}\right)\frac{\epsilon_F}{\hbar k_F}. 
\label{eq:alpha}
\end{equation}
with $k_{F}$ being the Fermi wave number and $\epsilon_{F}$ being the Fermi energy; {$(k_F l_{sk})^{-1}$ is given by the ratio of the SOC energy to the Fermi energy. We also assume that the exchange field $\bm{n}(\hat{\bm r} ,t)$ varies with the characteristic length given by $l_{sk}$. }In the following calculation, we consider the case when the coupling constant $J$ is large and the exchange field varies slowly on microscopic scales,
i.e., we assume the following conditions  
\begin{align}
&\left(\frac{\alpha_{so} \hbar k_F}{2JM}\right)=\left(\frac{\epsilon_{F}}{2JM}\right)\left(\frac{1}{k_{F} l_{sk}}\right)\ll 1,  \label{eq:condition1}\\
& \left(\frac{\epsilon_{F}}{2JM}\right)\left(\frac{\hbar}{\epsilon_{F} t_{sk}}\right) \ll 1, \label{eq:condition2}
\end{align}
where $t_{sk}$ is the time scale of the variation of the exchange field in real time, and $2 M J$ is the scale of the exchange splitting.
These conditions allow us to adopt an adiabatic approximation which assumes that the transitions between two bands are neglected. In the case without SOC, it corresponds to assuming that the spin will follow the direction of the exchange field locally in space and time. Here SOC field modifies the direction the spin follows depending on the momentum. 
  
Within the adiabatic approximation, we can construct a wave packet for each band.
We follow a previous work\cite{Sundaram1999} for the construction of wave packets and will not mention the detail here.
In the following semiclassical formalism, we calculate the current from the Boltzmann transport equation, and the effect of the impurity potential can be included as a relaxation time approximation.
Assuming that the spatial spread of the wave packet is much smaller than $l_{sk}$, we can expand the Hamiltonian around the wave packet center $\bm r_c$ as
\begin{align}
H&\thickapprox H_c+\Delta H,\\
H_c&=\frac{\hat{\bm p}^2}{2m}
-JM\bm{n}({\bm r_c} ,t)\cdot\bm{\sigma}
+\alpha_{so}\hat{\bm p}\cdot\bm{\sigma},\\
\Delta H&=-J M\left(\frac{ \partial \bm n(\bm r_c, t)}{\partial r_{\alpha}}\cdot\bm{\sigma}\right)(\hat {r_{\alpha}}-{r}_{c\alpha}),
\end{align} 
where we include the spatial variation of the exchange field as a perturbation $\Delta H$. 
Since the local Hamiltonian $H_c$ can be viewed as a homogeneous two level system, its eigenstates are written as
\begin{align}
| \psi_{\bm{k}+}(\bm r_c,t)\rangle&
=  e^{i \bm k\cdot \bm r_c}
\begin{pmatrix} \cos\frac{\Theta}{2}e^{-i\Phi}\\ \sin\frac{\Theta}{2}\end{pmatrix},
\label{eq:eigen1}\\
| \psi_{\bm{k}-}(\bm r_c,t)\rangle&=  e^{i \bm k\cdot \bm r_c}
\begin{pmatrix} \sin\frac{\Theta}{2}e^{-i\Phi}\\- \cos\frac{\Theta}{2}\end{pmatrix},
\label{eq:eigen2}
\end{align}
where $\Theta$ and $\Phi$ are defined by
\begin{align}
 \bm{  {h}}(\bm r_c,\bm k,t)& \equiv{ -J\bm M(\bm r_c, t)+\alpha_{so}\hbar \bm k},\\
 \bm{ \hat {h}}(\bm r_c,\bm k,t)&=\frac{\bm h(\bm r_c,\bm k,t)}{|\bm h(\bm r_c,\bm k,t)|}
\\
&\equiv(\sin\Theta\cos\Phi, \sin\Theta\sin\Phi, \cos\Theta).
\end{align}
Their eigenvalues are given by  $H_c| \psi_{\bm{k}\pm}(\bm r_c,t)\rangle=\left(\frac{\hbar^2 k^2}{2m}\pm |\bm h| \right)| \psi_{\bm{k}\pm}(\bm r_c,t)\rangle$.
Then using these eigenstates as a basis, we can construct wave packets, which follow the equations of motion:
\begin{align}
\dot {\bm{r}}_{\sigma}&=\frac{\partial\epsilon_{\sigma}}{\hbar\partial\bm k_{\sigma}}-{\Omega}^{kr}_{\sigma}\dot{\bm{r}}_{\sigma}
-\Omega^{kk}_{\sigma}\dot {\bm{k}_{\sigma}}-\bm{\Omega}^{kt}_{\sigma}, \label{eq:eom1}\\
\dot {\bm{k}}_{\sigma}&=-\frac{\partial\epsilon_{\sigma}}{\hbar\partial\bm r_{\sigma}}+{\Omega}^{rr}_{\sigma}\dot{\bm{r}_{\sigma}}
+\Omega^{rk}_{\sigma}\dot {\bm{k}_{\sigma}}+\bm{\Omega}^{rt}_{\sigma},\label{eq:eom2}
\end{align}
where $\sigma=\pm$ denotes band indices, $\bm r_\sigma$ is the wave packet center, $\bm k_\sigma$ is the mean wave vector of the wave packet, $\epsilon_\sigma$ is the wave packet energy, and a $3 \times 3$ antisymmetric matrix $\Omega$ and a $3$ component vector $\bm{\Omega}$ are the Berry curvatures defined in the parameter space $(\bm r, \bm k, t)$.

The Berry curvatures of the unperturbed states, Eqs. (\ref{eq:eigen1}) and (\ref{eq:eigen2}), are given by\cite{Freimuth2013, Xiao2010}
\begin{align}
\left (\Omega^{kr}_{\pm}\right)_{\alpha \beta}&=\mp \frac{1}{2}\hat{\bm h}\cdot
\left(\frac{\partial}{\partial k_\alpha}\hat{\bm h}\times \frac{\partial}{\partial r_\beta}\hat{\bm h}  \right), \label{eq:curvature1}\\ 
\left ( \Omega^{kk}_{\pm}\right)_{\ \alpha \beta}&=\mp \frac{1}{2}\hat{\bm h}\cdot
\left(\frac{\partial}{\partial k_\alpha}\hat{\bm h}\times \frac{\partial}{\partial k_\beta}\hat{\bm h}  \right), \label{eq:curvature2}\\
\left(\bm{\Omega}^{kt}_{\pm}\right)_{\ \alpha}&=\mp \frac{1}{2}\hat{\bm h}\cdot
\left(\frac{\partial}{\partial k_\alpha}\hat{\bm h}\times \frac{\partial}{\partial t}\hat{\bm h}  \right).
\label{eq:curvature3}
\end{align} 
$ \Omega^{rr}_{\sigma}, \Omega^{rk}_{\sigma}$ and $\bm{\Omega}^{rt}_{\sigma}$ are written in the same way (See Appendix\ref{Berry} for the definitions of Berry curvatures).

Here the perturbation $\Delta H$ introduces corrections to the energy and the eigenstates, thus the above Berry curvatures.  
Firstly, the wave packet energy is given by the band energy of $H_c$ and  a correction from $\Delta H$,\cite{Sundaram1999,Freimuth2013} which is written as 
\begin{align}
\epsilon_{\pm}&=\frac{\hbar k^2}{2 m}\pm |\bm h|+\delta \epsilon_{\sigma},\\
\delta \epsilon_{\sigma}&=\langle W_{\sigma}| \Delta H |W_{\sigma}\rangle=|\bm h |\mathrm{Tr}\  \Omega^{kr}_{-},
\end{align}
where $|W_{\sigma}\rangle$ is the wave packet constructed from the unperturbed state.
Note that the value of $\delta \epsilon_{\sigma}$ is the same in both bands. 
Secondary, $\Delta H$ gives the corrections to the above Berry curvatures, which are one order higher in $(k_F l_{sk})^{-1}$.\cite{Xiao2009} 
In the following, however, we will calculate the dissipationless current up to the order of $(k_F l_{sk})^{-2}$ or $(k_F l_{sk})^{-1}(\epsilon_F t_{sk}/\hbar)^{-1}$, and the unperturbed Berry curvatures are sufficient in this case. 

From now on, we will calculate the induced dissipationless current, which is independent of the relaxation time.
In a relaxation time approximation of the Boltzmann equation,  such currents are obtained from the equilibrium distribution function. 
Note that dissipative currents flow as a result of the deviation from the equilibrium distribution, and we will calculate them in the next section. 
Using the velocities of the wave packet center, the dissipationless current density is given by 
\begin{equation}
\bm j(\bm r, t)=e\sum_{\sigma}\int d \bm k D_\sigma (\bm k, \bm r)f(\epsilon_{\sigma})\dot{\bm r}_\sigma, 
\end{equation}
where $e=-|e|$ is the electron charge, $D_{\sigma}(\bm k, \bm x)$ is the density of states in phase space, $f(\epsilon)$ is the Fermi Dirac distribution function. In our calculation, the chemical potential is assumed to be constant in space and time. 
Berry curvatures modify the density of states from $(2\pi)^{-3}$ as \cite{Xiao2005}
\begin{align}
D_{\sigma}(\bm r, \bm k)=\frac{1}{(2\pi)^3}\sqrt{\mathrm{det}\left(\overline{\Omega}_{\sigma} -\overline{ J}\right)},
\end{align}
where $\overline{\Omega}_{\sigma}$ and $\overline{ J}$ are $6 \times 6$ matrices which are given by \cite{Xiao2005}
\begin{align}
\overline{\Omega}_{\sigma}&=
\begin{pmatrix} \Omega^{rr}_{\sigma}&  \Omega^{rk}_{\sigma}\\ \Omega^{kr}_{\sigma}& \Omega^{kk}_{\sigma}\end{pmatrix},\\
\overline{J}&=
\begin{pmatrix} 0&  I\\ -I& 0\end{pmatrix}.
\end{align}
Now we solve Eqs. (\ref{eq:eom1}) and (\ref{eq:eom2}) for $\bm{\dot{r}}_{\sigma}$, and calculate the current up to the order we consider.
The obtained current is described by
\begin{align}
\bm j (\bm r, t)=
-e\sum_{\sigma}\int \frac{ d \bm k}{(2\pi )^3}  f(\epsilon_{\sigma}^{(0)})\bm{\Omega}^{kt(2)}_{\sigma},
\end{align}
where
\begin{align}
\epsilon_{\pm}^{(0)}&=\frac{\hbar^2 k^2}{2m}\pm J M(\bm r, t), \\
\bm{\Omega}^{kt(2)}_{\pm}&=\pm \frac{\hbar \alpha_{so}}{2 JM}
\left(
\bm n(\bm r,t)\times \frac{\partial}{\partial t} \bm n(\bm r,t)
\right)
.\end{align}
The superscripts $(0)$ and $(2)$ denote the order of $(k_F l_{sk})^{-1}$  and $(\epsilon_F t_{sk}/\hbar)^{-1}$  to show that the Berry curvatures are expanded in the series of $\alpha$ noting the relation Eq. (\ref{eq:alpha}).
If we assume $(k_B T/\epsilon_F) \ll 1$ and $(JM/\epsilon_F)\ll1$ with $T$ being the temperature, we can perform the summation over the bands and obtain  
\begin{align}
\bm j (\bm r, t)\approx
N(\epsilon_F)e\hbar \alpha_{so}
\left(
\bm n(\bm r,t) \times \frac{\partial \bm n(\bm r,t)}{\partial t}
\right), \label{eq:adiabatic}
\end{align}
where $N(\epsilon)$ is the density of states defined by $N(\epsilon)=({m^{\frac{3}{2}}\epsilon^{\frac{1}{2}}})/({\sqrt2 \pi^2 \hbar^3})$.
Only the adiabatic current $\bm \Omega^{kt}$ remains within the order we consider, while other Berry curvatures does not contribute to the result. This adiabatic current is induced by the coupling between antisymmetric SOC and the time-dependent exchange field.

\section{Dissipative current}
In the last section, we derived the dissipationless part of the electric current by using the semiclassical theory: the equations of motion and the Boltzmann equation. From now we calculate the dissipative currents within the linear response against the temporal variance of the exchange field. Here we need to systematically expand the distribution function in the series of the derivatives, and thus it is more convenient to use the quasiclassical approach to the Green function rather than the Boltzmann equation.     
Our system in Eq. (\ref{eq:H}) can be described as the following Lagrangian in the second quantized form: 
\begin{align}
L&=L_{0}+H_{\rm{ex}}+H_{\rm{so}},  \label {eq:L}\\
L_{0}&=\int d\bm{x} c^{\dag}(x)\left( \hbar \frac{\partial}{\partial \tau}-\frac{\hbar^2}{2m}{\bm \nabla}^2 +v_{\rm{imp}}(\bm{r}) \right )c(x), \label{eq:zero}\\
H_{\rm{ex}}&=-JM\int d\bm{x} c^{\dag}(x)\bm{n}(x)\cdot\bm{\sigma} c(x), \label {eq:ex}\\
H_{\rm{so}}&=\alpha_{so} \int d\bm{x} c^{\dag}(x)\bm{g}\left(-i \hbar\bm\nabla\right)\cdot\bm{\sigma} c(x), \label {eq:so}
\end{align}
where $c^{\dag}(x)=(c^{\dag}_{\uparrow}(x),c^{\dag}_{\downarrow}(x))$ is the two-component spinor field operators for electrons with spin up ($\uparrow $) and down ($\downarrow$) along the $z$ axis, which depend on the four-vector $x=(\bm{r},\tau)$ with $\tau$ being imaginary time. 
As described in the last section, $H_{\rm{ex}}$ is the exchange interactions, $H_{\rm{so}}$ is the antisymmetric SOC with $\bm{g}_{so}(-i\hbar \bm{\nabla})=  (-i\hbar\partial_x, -i\hbar\partial_y, -i\hbar\partial_z)$.
In addition, the same conditions, Eqs. (\ref{eq:condition1}) and (\ref{eq:condition2}), are assumed. 
To simplify the calculation we first make the spin quantization axis oriented along $\bm{n}(x)$ locally\cite{Korenman1977, Tatara2008}, and define new spinor field operators $\psi^{\dag}(x)=\left(\psi^{\dag}_{+}(x), \psi^{\dag}_{-}(x) \right)$, which are written as
$c(x)=U(x) \psi(x)$ and $c^{\dag}(x)=\psi^{\dag}(x) U^{\dag}(x)$,
with a $2 \times 2$ unitary matrix 
\begin{equation}
U(x)=e^{-\frac{1}{2}i \phi\sigma_z}e^{-\frac{1}{2}i \theta\sigma_y}.
\end{equation}
 $U(x)$ transforms the exchange interactions term in Eq. (\ref{eq:ex}) as $U^{\dag}(x)\bm{n}(x)\cdot \bm{\sigma} U(x)=\sigma_z $. 
Thus,  $\psi^{\dag}_{+}(x)\  \left(\psi^{\dag}_{-}(x) \right) $ denotes the electrons antiparallel (parallel) to $\bm{n}(x)$.
We then obtain the Lagrangian written in terms of the transformed field $\psi(x)$ and $\psi^{\dag}(x)$ as 
\begin{align}
L_0&=\int d \bm{x}\psi^{\dag}(x)
          \left[ 
		\hbar \frac{\partial}{\partial \tau}
		+\hbar U^{\dag}(x)\frac{\partial}{\partial \tau} U(x)
	   \right.\nonumber\\
	   	&\hspace{10pt}\left.+\frac{1}{2m}
					\left(-i\hbar \bm\nabla
							-i\hbar  U^{\dag}(x)\bm{\nabla} U(x) \right)^2
		+v_{\rm{imp}}(\bm r)	   
   \right]\psi(x),		
		\\ 
H_{ex}&=-JM\int d\bm{x}\psi^\dag(x)\sigma_z\psi(x),\\
H_{so}&=	\alpha_{so}\int d\bm{x}\psi^\dag(x)\nonumber\\
&\hspace{30pt} \left(U^{\dag}(x)\bm{\sigma}U(x)\right)\cdot \left(-i\hbar \bm{\nabla}-i\hbar U^{\dag}(x)\nabla U(x) \right)
\psi(x). \label{eq:all}
\end{align}
Considering that the electrons tend to be parallel or antiparallel because of the large exchange coupling, we then neglect the off-diagonal term in the transformed Lagrangian, which causes the spin-flip transitions between majority and minority spin states.
Here we note that the diagonal part of $ i\hbar U^{\dag}(x)\nabla U(x)$ and 
$-\hbar U^{\dag}(x)\frac{\partial}{\partial \tau} U(x)$ can be written as 
$e\mathbf A ^{em}(x) \sigma_z$ 
and 
$e\mathrm{A}_0^{em}(x) \sigma_z $,
where $e<0$ is the electron charge, and
 $\mathbf{A}^{em}(x)$ and $\mathrm{A}_{0}^{em}(x)$ are defined by
\begin{align}
&\mathbf{A}^{em}(x)=\frac{\hbar}{2e} \cos{\theta}\bm{\nabla} \phi, \\
&\mathrm{A}^{em}_0(x)=i \frac{\hbar}{2 e}\cos{\theta} \partial_\tau \phi, 
\end{align}
respectively. 
Here $\mathbf{A}^{em}(x)$ and $\mathrm{A}^{em}_0(x)$ are the vector potential and scalar potential of the effective electromagnetic fields, so-called emergent electromagnetic fields. \cite{Volovik1987, Xiao 2010,Nagaosa2012} 
The emergent electromagnetic fields can be written as 
\begin{align}
E^{em}_{i}(x)&\equiv \left( -\frac{\partial}{\partial t}\mathbf{A}^{em}(x)+\bm \nabla \mathrm{A_0}^{em}(x)\right)_i
=-\frac{\hbar}{2 e}\bm n\cdot (\partial_i\bm n \times \partial_t \bm n), \label{eq:E}\\
B^{em}_{i}(x)&\equiv \left(\bm \nabla \times \mathbf{A}^{em}(x)\right)_i=-\frac{\hbar}{4e}\epsilon_{ijk}\bm n\cdot (\partial_j\bm n \times \partial_k \bm n), \label{eq:B}
\end{align}
in the real time representation $\tau \to i t$, where $\epsilon_{ijk}$ is the antisymmetric unit tensor and the subscript $(i,\ j,\ k )$ denotes the spatial coordinate $(x,\ y,\ z )$. 
These emergent fields are described as the nonzero Berry curvatures $\Omega^{rr}$ and $\bm{\Omega}^{rt}$, and they act on electrons in the same manner as the electromagnetic field.  
In Appendix\ref{emergent}, we demonstrate how $\mathbf{A}^{em}(x)$ and $\mathrm{A}^{em}_0(x)$ appear in a Lagrangian without SOC.
In our case, where SOC also gives rise to a force on electrons, 
we will treat it as the corrections to the emergent electromagnetic fields\cite{Kim2012a, Tatara2013} .

After the adiabatic approximation, the Lagrangian is given by
\begin{align}
L&=\int d \bm{x}\psi^{\dag}_{\sigma}(x)\nonumber\\
&\qquad \left[
		 \hbar \frac{\partial}{\partial \tau} \right. 
  		+\frac{1}{2m}\left(-i\hbar\bm{\nabla}
						-e q_\sigma \left(\mathbf{A}^{em}(x)
												-\frac{m \alpha_{so}}{e}\bm{n}(x)
												 \right)
						\right)^2	\nonumber\\
&\quad \qquad \left.
		-q_\sigma J M
		-e q_\sigma  \mathrm{A}^{em}_0(x)	
+v_{\rm{imp}}(\bm r)
+\mathrm{V}(x)
		\right]
\psi_{\sigma}(x), \label{eq:full}
\end{align}
where $\sigma=\pm$, $q_{+}=-1$ ($q_{-}=+1$) is attributed to the upper (lower) band, and $\mathrm{V}(x)/\epsilon_F$ is of the order of $(k_F l_{sk})^{-2}$ (See Appendix\ref{quasi}).
We will see that ${\mathrm{V}}(x)$ does not contribute to the final result.

For simplicity, we again introduce field operators with an additional phase written as 
$\psi_{\sigma}(x)=e^{-\frac{i}{2}\lambda(x) q_{\sigma}}\psi_{\sigma}'(x),$
where $\lambda(x)$ will be determined later.
Based on the above Lagrangian, we can define two independent Green functions $G_\sigma(x, x')=-\left<T_\tau \psi'_{\sigma}(x)\psi'^\dag_{\sigma}(x')\right>$, which satisfy the Dyson equations
\begin{align}
&\left[-\hbar \frac{\partial}{\partial \tau}-\zeta_\sigma \left(x,-i\hbar \bm{\nabla_r} \right)\right]G_\sigma(x,x')\nonumber\\
&\hspace{65pt}-\int d^4 y \Sigma_\sigma(x,y) G(y, x')=\delta(x-x').
\label{eq:dyson}
\end{align}
Here $\zeta_\sigma \left(x,-i\hbar \bm{\nabla_r} \right)$ is the single-particle energy operator, and 
$\Sigma_\sigma(x, y)$ is the self energy operator due to the impurity potential.

To investigate the local response of electrons at $(\bm{r}_0, \tau_0)$ against spatial and temporal variations 
of the exchange field, 
we consider an area within a radius of $ k_F^{-1}$ and $\hbar/ \epsilon_F$  from $\bm{r}_0$ and $\tau_0$, respectively.
By expanding slowly varying quantities around the point $(\bm r_0, \tau_0)$, we can treat such variations perturbatively. 
The single particle energy, in semiclassical form, can be expanded up to the order of $(k_F l_{sk})^{-2}$ or $(k_F l_{sk})^{-1}(\epsilon_F t_{sk}/\hbar)^{-1} $ as
\begin{align}
&\zeta_\sigma \left(x,\bm p \right)
=\frac{1}{2m}
	\left[ 
		\bm p
		-e q_\sigma 
						\bm{\mathcal{A}}^{\mathit{eff}}(x) 
	\right] ^2
	-\mu_{\sigma}
,\label{eq:energy}
\end{align}
where 
\begin{align}
&\bm{\mathcal{A}}^{\mathit{eff}}(x)=
												\left[\mathbf{A}^{em}(x_0)
												-\frac{m \alpha_{so}}{e}\bm{n}(x_0)
												 \right]	\nonumber
\\
												&\hspace{10pt}+\left[
														 \frac{\partial}{\partial r^\mu}\mathbf{A}^{em}(x_0)
														-\frac{m \alpha_{so}}{e}
															\frac{\partial}{\partial r^\mu}\bm{n}(x_0)
												\right] (r^\mu-r^\mu_0)	\nonumber
\\
												&\hspace{10pt}+\left[
														i \bm{\nabla} \mathrm{A}_0^{em}(x_0)
														+\frac{\partial}{\partial \tau}\mathbf{A}^{em}(x_0)
													       -\frac{m \alpha_{so}}{e}
														\frac{\partial}{\partial \tau}\bm{n}(x_0)
												\right] (\tau-\tau_0)
\label{eq: ene}
\end{align}

\begin{align}
\mu_\sigma = \epsilon_F +J M q_\sigma 
-\mathrm{V}(x_0)
\end{align}
and the subscript $\mu$ denotes the spatial coordinate $(\mu=x, y, z )$. 
In addition, we chose the form of $\lambda(x)$ so that the effective scalar potential is zero up to the order we consider. 
The explicit form of $\lambda(x)$ is written in Appendix \ref{quasi}.

In the above mentioned area around $(\bm r_0, \tau_0)$, the single-particle energy operator describes a charged particle moving in space-time independent electromagnetic fields, which are written as
\begin{align}
&\bm{\mathcal{E}}^{\mathit{eff}}(x_0)=\bm{E}^{em}(x_0)+\bm{E}^{so}(x_0),\label{eq:eff1}\\
&\bm{\mathcal{B}}^{\mathit{eff}}(x_0)=\bm{B}^{em}(x_0)+\bm{B}^{so}(x_0),\label{eq:eff2}
\end{align}
where the SOC-induced fields are given by
\begin{align}
\bm{E}^{so}(x)&=-\frac{m \alpha_{so}}{|e|}\frac{\partial}{\partial t}\bm{n}(x), \label{eq:electric}
\\
\bm{B}^{so}(x)&=\frac{m \alpha_{so}}{|e|}	\bm{\nabla}\times\bm{n}(x),\label{eq:magnetic}
\end{align}
in real time, and the emergent fields $\bm E^{em}(x)$ and $\bm B^{em}(x)$ are given by Eqs. (\ref{eq:E}) and (\ref{eq:B}). 
As we can see in Eq. (\ref{eq:energy}), the effective charge of the particle of each band is given by $e q_{\sigma=\pm}=\mp e$. $\bm{E}^{so}(x)$ and $\bm{B}^{so}(x)$ are effective electric and magnetic fields, which originate from the antisymmetric SOC, and they are of the same order in powers of  $(k_F l_{sk})^{-1}$ as the emergent electromagnetic fields. The emergent electric and magnetic fields are described by the Berry curvatures $\Omega_{rr}$ and $\bm{\Omega}_{rt}$ because they are due to the adiabatic motion of the spin, which follows the exchange field. On the other hand, the SOC-induced fields are not caused by the Berry curvatures; they are the response against the variance of the exchange field $\bm n(x)$ in the presence of the antisymmetric SOC.    

Here we comment on the SOC-induced electric field.
$\bm E ^{so}(x)$ can be viewed as a magnetoelectric effect\cite{Fujimoto2012} 
if we regard $\bm n(x)$ as an effective Zeeman field. The temporal variation of $\bm n(x)$ gives rise to the change in the spin distribution of conduction electrons, and it deforms the Fermi surface because of antisymmetric SOC, which couples the momentum and the spin of conduction electrons asymmetrically. 
Thus the term proportional to $\partial_t\bm{n}(x)$ act as an electric field.
  
We can now calculate the electric current by applying the quasiclassical transport equations derived by Houghton et al.\cite{Houghton1998}.  
This formalism allows us to derive the linear response to the effective electric field. 
The detail of the derivation is presented in Appendix\ref{quasi}, and here we present the final result.
The impurity scattering is treated using the Born approximation, 
and the quasiparticle lifetime $\tau_\mathit{imp}$ is introduced. 
Assuming $\tau_\mathit{imp}/t_{sk}\ll1$ and $MJ/\epsilon_F \ll 1$, we obtain the electric current
\begin{align}
\bm j(x)&=
N(\epsilon_F)\left(\frac{2e^2JM}{m}\right)
\tau_\mathit{imp}
\bm{\mathcal{E}}^{\mathit{eff}}(x)
\nonumber\\
&\hspace{40pt}
-N(\epsilon_F)\left(\frac{4e^3 \epsilon_F}{3m^2}\right) 
\tau_\mathit{imp}^2
\bm{\mathcal{B}}^{\mathit{eff}}(x)
\times
\bm{\mathcal{E}}^{\mathit{eff}}(x),
\label{eq:current1}
\end{align}
with $x=(\bm r, t)$.
The first part represents the longitudinal current, while the second part the Hall current.
Although electrons in the upper and lower bands have the opposite charge, the lower band electrons have the larger Fermi volume, which means that their contribution remain in the longitudinal current.
\section{Evaluations --Skyrmions, monopole and emergent dyon}
In this section, we evaluate the electric current and the effective electromagnetic fields in a skyrmions-merging process.
First of all, let us explain how this process occurs in Fe$_{0.5}$Co$_{0.5}$Si. For the detail of the setup, we refer to the paper of Milde et al \cite{Milde2013}. 
In Fe$_{0.5}$Co$_{0.5}$Si and MnSi, skyrmions form a two-dimensional triangular lattice which is translationally invariant in the direction parallel to an applied magnetic field (See Fig. \ref{fig:up} (a) for the illustration). 
In a bulk crystal, such a skyrmion lattice is observed only in a narrow region of the temperature and magnetic field phase diagram, which locates just below the critical temperature of the ordered phase.
When the skyrmion lattice is cooled keeping the applied field unchanged, however, it survives as a metastable state over a wide temperature range. Then, well below the critical temperature,
by decreasing the magnetic field, they observed that neighboring skyrmions coalesce into one elongated skyrmion; the number of skyrmions is reduced. The driving force of the dynamics originates from the energy difference between the metastable state and the lower energy state without skyrmions.  

Our study focuses on dynamics of two lines of skyrmions merge, which was observed in their numerical simulation inside a bulk crystal.
{We consider a three-dimensional model of $\bm n (\bm r, t)$ which varies in space and time so that it describes a monopole moves along the positive $z$ direction with a constant velocity.} (See Fig. {\ref{fig:up}}(b) for the illustration) 

Here the skyrmion helicity, the spin swirling direction, is chosen so that it corresponds to the case of $\alpha_{so}>0$; they are related by the sign of the DM interaction. We will only present the result for $\alpha_{so}>0$ here, and comment on the case of $\alpha_{so}<0$ later.    
Our model describes that a hedgehog monopole moves at a constant speed in the positive $z$ direction, and two lines of skyrmions coalesce into one skyrmion with a larger radius.  
Here two skyrmions get closer as the monopole approaches, and after the monopole passes through the single skyrmion does not vary in time.  
Further detail of this model is presented in Appendix\ref{model}.

Here we comment on the topological aspect. \cite{Milde2013}
The integral of the emergent magnetic field (Eq. (\ref{eq:B})), over a closed surface is proportional to the topological invariant, the number of times the map $\bm n (x)$ covers the unit sphere.
By identifying the boundary of an area, an area can be compactified to a sphere, closed surface.
A hedgehog monopole and a skyrmion are characterized by this topological number. 
In our model, a $xy$ plane above the monopole gives the topological number $-2$, which indicates that two quanta of emergent magnetic flux flows in the negative $z$ direction. On the other hand, a $xy$ plane below the monopole gives the topological number $-1$.
Thus, the topological number of a sphere surrounding the monopole is $-1$; it is an magnetic anti-monopole of the emergent field because it absorbs one quantum of emergent magnetic flux.
\begin{figure}
\begin{center}
\includegraphics[bb=10 150 276 500 clip, width=0.95\columnwidth]{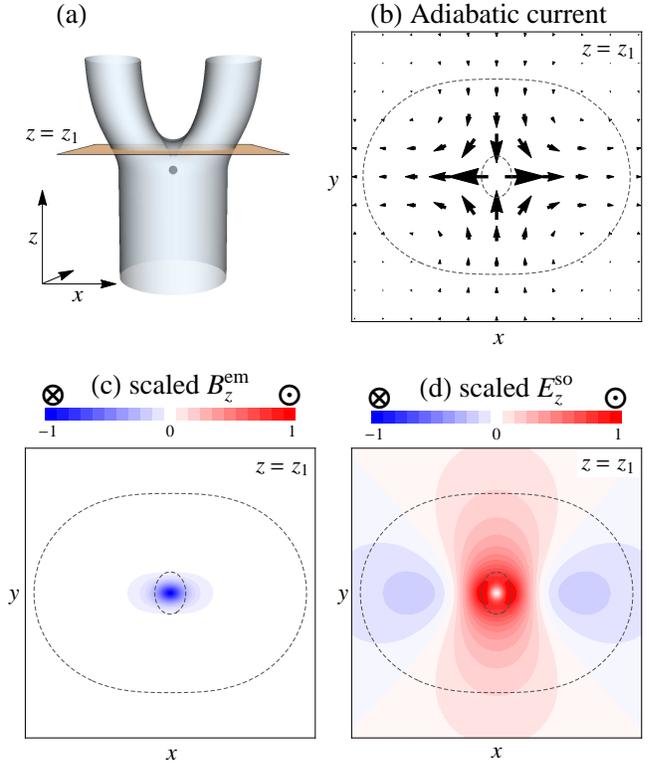}
\end{center}
		\caption{(Color online)
(a) A $xy$ plane at $z=z_1$, just above the monopole. The black dot indicates the position of the monopole, which moves in the positive $z$ direction with a constant velocity. The current and the fields are evaluated in this plane.     
(b) The flow of the in-plane dissipationless current, the adiabatic current given by Eq. (\ref{eq:adiabatic}).
The sizes and the directions of the arrows indicate the density and the direction of the current respectively.  
The dashed line indicates the cross section of the curved surface in (a), which characterizes the profile of skyrmions.     
(c) The distribution of the $z$ component of the emergent magnetic field. The field strength is scaled by the maximum of the absolute value. 
(d) The distribution of the $z$ component of the SOC-induced electric field.
}\label{fig:panel}
\end{figure}
From now, to investigate the electrodynamics induced by a moving monopole, we consider a plane right above the point defect, which is described in {Fig. \ref{fig:panel} (a)}, at a certain time $t=t_1$. 
We define the $z$ coordinate of the plane as $z=z_1$. The distance of the plane and the point defect is nearly equal to the radius of a single skyrmion.   
First of all, we show the in-plane flow of the dissipationless current, the adiabatic current (Eq. (\ref{eq:adiabatic})) in {Fig. \ref{fig:panel} (b)}. 
It is driven by the motion of the exchange field when the centers of the skyrmions move. 
Approximately, the direction of the current is the opposite to the velocity of a skyrmion in the $x$ direction.

Then we calculate the distribution of the effective field due to the the dynamics, which drive the dissipative currents.
There are two contributions in the effective field in Eqs. (\ref{eq:eff1}) and (\ref{eq:eff2}); the emergent electromagnetic field and the field due to SOC. 
Let us first consider the emergent electromagnetic field. 
As mentioned above, the hedgehog monopole has a quantized magnetic charge. 
{Fig. \ref{fig:panel} (c)} describes the concentration of the magnetic flux, which flows into the monopole. Note that the monopole emits the  half of the incoming flux in negative $z$ direction. Since the flux is confined in skyrmions, the distribution of the magnetic field around the monopole is anisotropic. 
In addition, from Eqs. (\ref{eq:E}) and (\ref{eq:B}) one obtains the relation
\begin{equation}
\bm \nabla \times \bm E^\mathit{em}=-\frac{\partial \bm B^\mathit{em}}{\partial t},
\end{equation}
which is the same as the Maxwell-Faraday equation. 
Thus the circular electric field occurs in a $xy$ plane as the magnetic monopole moves in the $z$ direction.

Next we consider the SOC-induced electromagnetic field, particularly focus on the electric field, which is proportional to $-\partial_t \bm n$. 
The $z$ component of the electric field has a nonzero value as shown in {Fig. \ref{fig:panel} (d)}, while the $x$ and $y$ components are zero on average in a $xy$ plane.
Furthermore, the electric field does not exist below the monopole since we assume $\bm n(x)$ is time-independent after the monopole passes through.
Thus moving monopole can be viewed as emitting nonzero electric field, and we can regard the moving monopole as having electric charge of the SOC-induced electric field. Note that this feature can not be obtained in terms of the the emergent electric field, Eq. (\ref{eq:E}). 
This is the main result of this paper. The hedgehog monopole can be viewed as having both the electric charge and the magnetic charge in terms of the effective electromagnetic field in Eqs. (\ref{eq:eff1}) and (\ref{eq:eff2}). 
In other words, the moving monopole at the merging point of skyrmions behaves like a dyon for conduction electrons. 

We here discuss on the above result that a monopole hl{acts} as a dyon-like particle.
Firstly, to obtain the result, we assume that the single skyrmion after coalescing is larger than one before coalescing, which leads that the time derivative of $n_z(x)$ is negative on average around the merging point. 
In experiments an elongated skyrmion was observed after coalescence, and thus the assumption is not model specific. 
{
Secondly, we comment on the effective electric charge density, defined by $\rho^{\mathit{eff}}=\bm{\nabla}\cdot\bm{\mathcal{E}}^{\mathit{eff}}$. As shown in Fig. 3, the positive charge is concentrated around the magnetic charge, while far from the point defect, lower density of negative charge spreads as if it screens the positive charge around the defect. In this system, the total effective charge is zero. Nevertheless, since electrons close to the monopole feel nonzero net electric fields generated by the charge concentrated in the vicinity of the monopole, as shown in Fig.3, the monopole behaves like a dyon for nearby electrons. }
{Thirdly,} in the above, we considered a hedgehog monopole with the topological number $-1$, which has negative magnetic charge and positive electric charge. 
On the other hand, a hedgehog monopole with the topological number $+1$ is generated at the same time. \cite{Milde2013}. 
In this case, the monopole has positive magnetic charge and negative electric charge.
Finally, the effective electric charge in a monopole in this paper is distinguished from the charge in a skyrmion studied by Freimuth \textit{et al.}\cite{Freimuth2013} We define the electric charge using the effective electric field induced by the dynamical effect, while Freimuth \textit{et al.} argued the concentration of the electron density due to the static effect. 

In the merging process, the SOC-induced electric field is considered to be weaker than the emergent electric field, though they are of the same order in $({k_F l_{sk}})^{-1}$. This is because the spatial variation is larger than $l_{sk}^{-1}$ around the merging point of skyrmions. 
{Nevertheless, the directional dependence may allow us to detect the SOC-induced field by measuring the longitudinal current, which is proportional to $\mathcal{\bm{E}}^\mathit{{eff}}$.  The SOC-induced electric field is parallel to the positive $z$ direction, while the emergent electric field has the in-plane direction, which is circular in a $xy$ plane. Thus, the detection of the longitudinal current in the $z$ direction is the way to observe the SOC-induced electric field.  
However, the existence of the Hall current, which is proportional to $\bm{\mathcal{B}}^{\mathit{eff}}\times \bm{\mathcal{E}}^{\mathit{eff}}$, renders the problem subtler. The Hall current flows along the $z$ axis, while there is no net current flowing in a $xy$ plane. Here, even without SOC, the Hall current in the $z$ direction flows; i.e. ${\bm{B}}^{em}\times {\bm{E}}^{em}$ also contributes to it. In order to discern the longitudinal current from the Hall current and detect the SOC-induced electric field, we can use their difference in the dependence on the relaxation time as shown in Eq. ({\ref{eq:current1}}). We also note that the electric currents are detectable when the monopole moves through the surface of a sample attached to a lead, since the flowing current charges are almost confined in a finite region around the moving monopole. }
\begin{figure}[t]
		\begin{center}
			\includegraphics[bb=  30 40 380 300 clip, width=0.8\columnwidth]{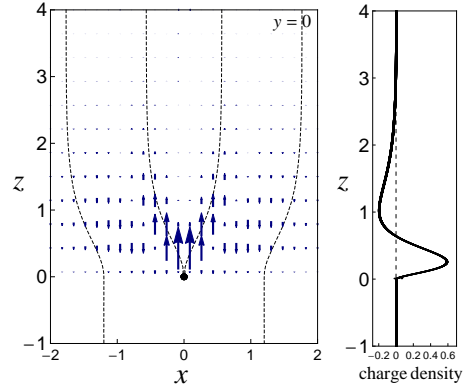}
		\end{center}
		\caption{{(Color online) (left) Distribution of the SOC-induced electric field at the cross section $y=0$. The dashed line indicates the profile of skyrmions projected on the $xz$ plane, and blue arrows show the intensity and the direction of the SOC-induced electric field. The black dot shows the position of the point defect, where the magnetic charge exists, and it can be viewed to emit the electric field. (right) Density of the effective charge per unit height in the $z$ direction.  It is obtained by integrating the charge density over a $xy$ plane, and scaled values are indicated.  The positive electric charge concentrates around the magnetic charge. Note that the emergent electric field does not contribute to the charge density.}}
 \label{fig:charge}
\end{figure}

So far we have shown the result for $\alpha_{so}>0$. 
On the other hand, for $\alpha_{so}<0$ we consider the skyrmions with the opposite helicity because of the opposite sign of the DM interaction.    
In this case, the induced in-plane adiabatic current is the same, but the $z$ component of the SOC-induced electric field has the opposite sign. Thus the sign of the electric charge of a monopole is the opposite, while the magnetic charge is the same because the emergent electromagnetic fields are independent of the SOC.  
The sign of a DM interaction is known to be changed by the chirality of the crystal, which can be different in crystal domains.\cite{Morikawa2013} 
This fact is supposed to give the variation of the property of a 'dyon'.

\section{Summary}
In this paper, we have investigated the electric currents and the effective electromagnetic field induced by the adiabatic dynamics of two skyrmions merging in a chiral magnet. 
By taking into account the antisymmetric SOC, we have obtained two types of contributions to the electric current: a dissipationless current; and a dissipative currents driven by the effective electromagnetic fields. 
The flow of the dissipationless current, an adiabatic current, and the spatial distribution of the effective field were evaluated in the above dynamics, and in terms of the effective electromagnetic fields the monopole at the merging point of two skyrmions turned out to be a dyon-like object. 
Our work has pointed out novel properties due to the interplay between the antisymmetric SOC and the dynamics of topological objects, skyrmions and a monopole.  
We expect that such properties also provide the controllability of the topological objects. 

\section*{Acknowledgments}
RT would like to thank M. Nitta for invaluable discussions. 
We also thank A. Shitade for the helpful comments on the manuscript. 
This work was supported by the Grant-in-Aids for Scientific Research from MEXT of Japan [Grants No. 23540406, No. 25103714 (KAKENHI on Innovative Areas ”Topological Quantum Phenomena”) and No. 25220711].

\onecolumn
\appendix
\section{Berry curvatures and wave packets}
\label{Berry}
In Appendix\ref{Berry}, we indicate the definitions of the Berry curvatures following Xiao \textit{et. al}.\cite{Xiao2010}.
Firstly, we derive the Berry curvatures for unperturbed states, Eqs. (\ref{eq:eigen1}) and (\ref{eq:eigen2}).
We rewrite the eigenstates as
\begin{align}
| \psi_{\bm{k}\sigma}(\bm r,t)\rangle&\equiv e^{i \bm k\cdot \bm r_c}|u_{\sigma}\rangle,
\end{align}
where $\sigma=\pm$ denotes the band index and $|u_{\sigma}\rangle$ depends on $(\bm k, \bm r, t )$. 
Their Berry connections $\mathcal A$ are defined by  
\begin{align}
\bm{\mathcal A}^{k}_{\sigma}=\langle u_{\sigma}|\frac{\partial}{\partial \bm k}|u_{\sigma}\rangle,\\
\bm{\mathcal A}^{r}_{\sigma}=\langle u_{\sigma}|\frac{\partial}{\partial \bm r}|u_{\sigma}\rangle,\\
\mathcal A^{t}_{\sigma}=\langle u_{\sigma}|\frac{\partial}{\partial t}|u_{\sigma}\rangle,
\end{align}
Then the Berry curvatures are given by, for example,  
\begin{align}
\left(\Omega^{kr}\right)_{\alpha\beta}
&=\frac{\partial}{\partial k_{\alpha}}\mathcal A^{r}_{\beta}-\frac{\partial}{\partial r_{\beta}}\mathcal A^{r}_{\alpha},\\
\left(\Omega^{kk}\right)_{\alpha\beta}
&=\frac{\partial}{\partial k_{\alpha}}\mathcal A^{k}_{\beta}-\frac{\partial}{\partial k_{\beta}}\mathcal A^{k}_{\alpha},\\
\left(\Omega^{kt}\right)_{\alpha}
&=\frac{\partial}{\partial k_{\alpha}}\mathcal A^{t}-\frac{\partial}{\partial t}\mathcal A^{k}_{\alpha}.
\end{align}
$\Omega^{rr}, \Omega^{rk}$ and $\bm{ \Omega}^{kt}$ are defined in the same way.
The straightforward calculations lead to Eqs. (\ref{eq:curvature1}) $-$ (\ref{eq:curvature3}).

\section{Emergent electromagnetic fields}
\label{emergent}
Here, we show emergent electromagnetic fields without SOC.\cite{Volovik1987, Xiao2010}
We consider a Lagrangian $L'=L_0 -H_{\rm{ex}}$, which is written as
\begin{equation}
L'=\int d\bm{x} c^{\dag}(x)\left( \hbar \frac{\partial}{\partial \tau}- \frac{\hbar^2}{2m}{\bm \nabla}^2
-J\bm{M}(x)\cdot\bm{\sigma} +v_{\mathit{imp}}(\bm{x}) \right )c(x),
\end{equation} 
by using Eqs. (\ref{eq:zero}) and (\ref{eq:ex}). 
We then rewrite the Lagrangian using $\psi(x)=U^{\dag}(x)c(x)$ as 
\begin{align}
L'
&=\int d \bm{x}\psi^{\dag}(x)
		 \left[
				 \hbar \frac{\partial}{\partial \tau}
		  		+\frac{1}{2m}\left(-i\hbar\bm{\nabla}
						-i\hbar U^{\dag}(x)\bm \nabla U(x)
						\right)^2	
		- J M \sigma_z
		+\hbar U^{\dag}(x)\frac{\partial}{\partial \tau}U(x)
		+v_{\mathit{imp}}(x)
		\right]
\psi(x).
\end{align}
We neglect the off-diagonal term which causes spin-flip transitions considering the exchange splitting is large. 
Thus the Lagrangian is 
\begin{align}
L'&=\int d \bm{x}\psi^{\dag}(x)
		 \left[
				 \hbar \frac{\partial}{\partial \tau}
		  		+\frac{1}{2m}\left(-i\hbar\bm{\nabla}
						-e \mathbf{A}^{em}(x)\sigma_z									
						\right)^2	
		- J M \sigma_z
		-e \mathrm{A}^{em}_0(x)	\sigma_z
+v_{\mathit{imp}}(x)
+\mathrm{V}'(x)
		\right]
\psi(x),\\
&=\int d \bm{x}\psi^{\dag}_{\sigma}(x)
		 \left[
				 \hbar \frac{\partial}{\partial \tau}
		  		+\frac{1}{2m}\left(-i\hbar\bm{\nabla}
						-e q_\sigma \mathbf{A}^{em}(x)									
						\right)^2	
		-q_\sigma J M
		-e q_\sigma  \mathrm{A}^{em}_0(x)	
+v_{\mathit{imp}}(x)
+\mathrm{V}'(x)
		\right]
\psi_{\sigma}(x),\label{eq:emergent}
\end{align}
where the potential 
$\mathrm{V}'(x)=\frac{\hbar^2}{8m} \left( (\bm{\nabla}\theta)^2+(\bm{\nabla}\phi)^2 \right)$ 
results from  $\left(i\hbar U^{\dag}\bm{\nabla}U \right)^2$.
From Eq. (\ref{eq:emergent}) we can interpret $\mathbf A^{em}$ and  $\mathrm{A}^{em}_0(x)$ 
as the vector and scalar potential of the emergent electromagnetic fields given 
by Eqs. (\ref{eq:E}) and (\ref{eq:B}).  

\appendix
\section{Quasiclassical calculations for the dissipative currents}
\label{quasi}
In Appendix B, we give the detail of the calculation in Sect. 3.
We begin with the Lagrangian in Eq. (\ref{eq:full})
\begin{align}
L=-\int d \bm{x}\psi^{\dag}_{\sigma}(x)
            \left[
		 \hbar \frac{\partial}{\partial \tau} 
  		+\frac{1}{2m}\left(-i\hbar\bm{\nabla}
						-e q_\sigma \left(\mathbf{A}^{em}(x)
												-\frac{m \alpha_{so}}{e}\bm{n}(x)
												 \right)
						\right)^2	
		-q_\sigma J M
		-e q_\sigma  \mathrm{A}^{em}_0(x)	
+v_{\mathit{imp}}(x)
+\mathrm{V}(x)
		\right]
\psi_{\sigma}(x),
\end{align}
where the potential $\mathrm{V}(x)$ is given by
\begin{equation}
\mathrm{V}(x)=\left[-\frac{m}{2}\alpha_{so}^2  + e\alpha_{so}\bm{n}(x)\cdot\mathbf{A}^{em}(x)
			 +\alpha_{so} \frac{\hbar}{2}\left( 
														\sin\phi\partial_x\theta -\cos\phi 	
														\partial_y\theta-\partial_z\phi	
												\right)
			+\frac{\hbar^2}{8m}\left( (\bm{\nabla} \theta)^2 +(\bm{\nabla} \phi)^2    \right)
			\right].
\end{equation}
We then express the Lagrangian using $\psi_{\sigma}'(x)=e^{\frac{i}{2}\lambda(x) q_{\sigma}}\psi_{\sigma}(x)$ as
\begin{align}
L&=\int d \bm{x}{\psi'}^{\dag}_{\sigma}(x)
            \left[
		 \hbar \frac{\partial}{\partial \tau} 
  		+\frac{1}{2m}\left(-i\hbar\bm{\nabla}
						-e q_\sigma \left(\mathbf{A}^{em}(x)
												+\frac{\hbar}{2e}\bm{\nabla}\lambda(x)
												-\frac{m \alpha_{so}}{e}\bm{n}(x)
												 \right)
						\right)^2	\right. \nonumber\\ 
&\hspace{100pt}\left.
		-q_\sigma J M
		-e q_\sigma\left(  \mathrm{A}^{em}_0(x)+i\frac{\hbar}{2e}\frac{\partial}{\partial \tau}\lambda(x)	\right)
+v_{\mathit{imp}}(x)
+\mathrm{V}(x)
		\right]
\psi_{\sigma}'(x).
\end{align}
Based on the Lagrangian, we write the Dyson equation for 
$G_{\sigma}(x,x')=-\left< T_\tau \psi_{\sigma}(x)\psi^\dag_{\sigma}(x')  \right> $,
\begin{equation}
\left[-\hbar \frac{\partial}{\partial \tau}-\zeta_\sigma \left(x,-i\hbar \bm{\nabla_x} \right)\right]G_\sigma(x,x')
-\int d^4 y \Sigma(x,y)_{\sigma} G_\sigma(y, x')=\delta(x-x'),
\end{equation}
where the one particle energy is given by
\begin{align}
\zeta_\sigma \left(x,-i\hbar \bm{\nabla} \right)
&=\frac{1}{2m}\left(-i\hbar\bm{\nabla}
						-e q_\sigma \left(\mathbf{A}^{em}(x)
												+\frac{\hbar}{2e}\bm{\nabla}\lambda(x)
												-\frac{m \alpha_{so}}{e}\bm{n}(x)
												 \right)
						\right)^2	
		-q_\sigma J M
		-e q_\sigma\left(  \mathrm{A}^{em}_0(x)+i\frac{\hbar}{2e}\frac{\partial}{\partial \tau}\lambda(x)	\right)
+v_{\mathit{imp}}(x) +\mathrm{V}(x) -\epsilon_F 
\end{align}
In the following we do not the take summation over repeated $\sigma$. 
To investigate the local response to the variations of the exchange field, 
we expand $\zeta_\sigma \left(x,-i\hbar \bm{\nabla_x} \right)$ around a point $x_0$ as described in Sect. 2.
Here we choose $\lambda$ so that it satisfies 
\begin{align}
i\frac{\hbar}{2e}\frac{\partial}{\partial \tau} \lambda(x)
 &=-\mathrm{A}^{em}_0(x_0)-\bm{\nabla}\mathrm{A}^{em}_0(x_0)\cdot (\bm{x}-\bm{x}_0)
	-\frac{\partial}{\partial \tau}  \mathrm{A}^{em}_0(x_0)(\tau-\tau_0), 
\\
\frac{\hbar}{2e}\bm{\nabla} \lambda(x)
 &= i \bm{\nabla}\mathrm{A}^{em}_0(x_0)(\tau-\tau_0).
\end{align}
We then obtain the one particle energy in Eq. (\ref{eq:energy}). 
From these Dyson equations, we calculate the electric currents using quasiclassical transport equations derived 
by Houghton et al.\cite{Houghton1998}
For the detail of the formalism we refer to Houghton et al., and we here present how to apply the formalism to our case.
We introduce the Wigner transformation of the Green function,
\begin{align}
G_{\sigma}(\bm p,\bm R; \tau, \tau')\equiv\int d \bm r G_{\sigma}\left(\bm R+\frac{\bm r}{2},\bm R-\frac{\bm r}{2}; \tau, \tau' \right)
e^{-i \frac{\bm p}{\hbar}\cdot \bm r},
\end{align}
where $\bm R = (\bm x +\bm x')/2$ and $\bm r =\bm x-\bm x' $ are center of mass and relative coordinates, respectively. 
For a local operator we introduce the circle product, which is written by
\begin{eqnarray}
\int d \bm r 
\zeta_{\sigma} (-i\hbar \bm\nabla_{\bm x},\bm{x},\tau)G_{\sigma}( x,  x')
e^{-\frac{i}{\hbar} \bm p\cdot \bm r}
&=&
\exp\left[\frac{i \hbar}{2}(\bm{\nabla}_{\bm{p}_2}\bm{\nabla}_{\bm{R}_1}-\bm{\nabla}_{\bm{p}_1}\bm{\nabla}_{\bm{R}_2})
\right] \zeta_{\sigma}(\bm{p}_1,\bm{R}_1,\tau)\left. G(\bm{p}_2,\bm{R}_2,\tau, \tau')\right|_{\bm{p}_1=\bm{p}_2=\bm{p}'}^{ \bm{R}_1=\bm{R}_2=\bm{R}}\nonumber\\
&\equiv&
\zeta_{\sigma}\left (\bm p, \bm R,\tau\right)
\circ
G_{\sigma}(\bm p, \bm R; \tau, \tau'),\\
\int d \bm r\int d  y
\Sigma_{\sigma}(x, y)G_{\sigma}( y,  x')e^{-\frac{i}{\hbar} \bm p\cdot \bm r}
&=&
\int d\tau_1 
\exp\left[\frac{i \hbar}{2}(\bm{\nabla}_{\bm{p}_2}\bm{\nabla}_{\bm{R}_1}-\bm{\nabla}_{\bm{p}_1}\bm{\nabla}_{\bm{R}_2})
\right] \Sigma_{\sigma}(\bm{p}_1,\bm{R}_1,\tau)\left. G(\bm{p}_2,\bm{R}_2,\tau, \tau')\right|_{\bm{p}_1=\bm{p}_2=\bm{p}'}^{ \bm{R}_1=\bm{R}_2=\bm{R}} \nonumber\\
&\equiv&
\int d\tau_1 \Sigma_\sigma (\bm p, \bm R;\tau, \tau_1)\circ G_{\sigma}(\bm p, \bm R; \tau_1, \tau').
\end{eqnarray}
Using the above relation we can write the Dyson equation as 
\begin{align}
-\frac{\partial}{\partial \tau}G_{\sigma}(\bm p, \bm R; \tau, \tau')
-\zeta_{\sigma}(\bm p, \bm R,\tau)\circ G_{\sigma}(\bm p, \bm R; \tau, \tau')
-\int d\tau_1 \Sigma_\sigma (\bm p, \bm R;\tau, \tau_1)\circ G_{\sigma}(\bm p, \bm R; \tau_1, \tau')
=\delta(\tau-\tau').\label{eq:right}
\end{align}
We can also write the Dyson equation in the form 
\begin{align}
\frac{\partial}{\partial \tau'}G_{\sigma}(\bm p, \bm R; \tau, \tau')
- G_{\sigma}(\bm p, \bm R; \tau, \tau')\circ\zeta_{\sigma}(\bm p, \bm R,\tau')
-\int d\tau_1  G_{\sigma}(\bm p, \bm R; \tau, \tau_1)\circ\Sigma_\sigma (\bm p, \bm R;\tau_1, \tau')
=\delta(\tau-\tau'),\label{eq:left}
\end{align}
which has the same physical information as Eq. (\ref{eq:right}).
The single particle energy in Eq. (\ref{eq:energy}) can be rewritten up to the second order of  as  
\begin{align}
\zeta_{\sigma}(\bm p, \bm R,\tau)
=\frac{1}{2m}(\bm p -eq_\sigma(\bm a+\bm u(\bm R)+\bm T(\tau)) )^2 -\mu_\sigma, 
\end{align}
where
\begin{align}
\bm a &\equiv \mathbf{A}^{em}(x_0)-\frac{m \alpha_{so}}{e}\bm{n}(x_0),\\
\bm u(\bm R) &\equiv \left[\frac{\partial}{\partial x^\mu}\mathbf{A}^{em}(x_0)
														-\frac{m \alpha_{so}}{e}
															\frac{\partial}{\partial x^\mu}\bm{n}(x_0)\right]
												 (R^\mu-x^\mu_0),\\
\bm T(\tau)&\equiv \left[i \bm{\nabla} \mathrm{A}_0^{em}(x_0)
								+\frac{\partial}{\partial \tau}\mathbf{A}^{em}(x_0)
								 -\frac{m \alpha_{so}}{e}
								\frac{\partial}{\partial \tau}\bm{n}(x_0)
								\right] (\tau-\tau_0),\\
				&\approx \bm T_m e^{-i\omega_m (\tau-\tau_0)},				
\end{align}
with $\mu = x, y ,z $.
Here, to simplify the calculation, we introduced $\bm T_m$, and $\omega_m$ which is of the order of $1/\tau_{sk}$.
We now subtract Eq. (\ref{eq:left}) from Eq. (\ref{eq:right}), integrate it over the quasiparticle energy $\zeta_{\sigma}$,
 and expand in powers of the spatial gradient. Then the obtained equation is written as
\begin{align}
&\left[ i \bm v_{\sigma} (s)\cdot \bm{\nabla}_{\bm R} + 
i e q_{\sigma} v_{\sigma\alpha}(s)\left (\frac{\partial u_\alpha(\bm R)}{\partial{ R_{\beta}}}-
\frac{\partial u_\beta(\bm R)}{\partial{ R_{\alpha}}}
\right) 
\frac{\partial s_i}{\partial p_\beta}
\frac{\partial }{\partial s_i}
+i\omega_n-i\omega_n'
\right]g_{\sigma}(s, \bm R; \omega_n, \omega_n') -T\sum_{\omega_k}[\sigma_{\sigma}(s,\bm R; \omega_n, \omega_k)g_{\sigma}(s, \bm R; \omega_k, \omega_n')\nonumber\\
&\hspace{60pt}-g_{\sigma}(s, \bm R; \omega_n, \omega_k)\sigma_{\sigma}(s,\bm R; \omega_k, \omega_n')
]
+e q_\sigma\bm v_{\sigma}(s)\cdot\bm T_m [g_{\sigma}(s, \bm R; \omega_n-\omega_m, \omega_n')
-g_{\sigma}(s, \bm R; \omega_n, \omega_n'+\omega_m)]
=0.
\end{align} 
Here we change the variables $(\bm p , \bm R)$ to $(s, \zeta_\sigma, \bm R)$ with $s$ being a parametrization of the Fermi surface, and introduce a quasiclassical propagator
 \begin{align}
g_{\sigma}(s, \bm R;\tau, \tau')&\equiv\frac{1}{\pi}\int d\zeta G_{\sigma}(\bm p, \bm R; \tau,\tau').
\end{align}
In addition, we transform it to the fermionic frequencies $\omega_n =(2n+1)\pi T$,
\begin{align}
g_{\sigma}(s, \bm R;\tau, \tau')&=T^2\sum_{n,n'} g_{\sigma}(s,\bm R; \omega_n,\omega_n')\exp(-i\omega_n (\tau-\tau_0) +i \omega_n' (\tau'-\tau_0)).
\end{align} 
The Fermi velocity $\bm v(s)$ is defined as 
\begin{align}
\left.\frac{\partial \zeta_\sigma}{\partial \bm p}\right|_{\bm p= \bm p_F}
\approx \sqrt{\frac{2(\epsilon_F +JMq_\sigma)}{m}} \frac{\bm p_F-eq_\sigma \bm a}{|\bm p_F-eq_\sigma \bm a|}
\equiv\bm v_{\sigma}(s), \nonumber\\
\end{align}
and in the following we assume  a spherical Fermi surface; 
$\bm v _{\sigma}\equiv v_{\sigma} (\sin \Theta \cos \Phi, \sin \Theta \sin\Phi ,\cos \Theta)$ 
with $\Theta$ and $\Phi$ being the parameterizations of the Fermi surface.  
The self energy is approximated  by the value at $|\bm{p}|=p_F$, and defined by $\sigma_{\sigma}(s, \bm R ; \omega_n, \omega_n')$. 
We here use the Born approximation to deal with the isotropic impurity scattering, 
thus the self energy is written as 
\begin{equation}
\sigma_{\sigma}(\omega_n)=\frac{1}{2\tau_\mathit{imp}} \int d s^2 g_{\sigma},
\end{equation}
where $\tau_{imp}$ is the quasiparticle lifetime.

To calculate the electric current within the linear response, we separate the integrated Green function as 
\begin{equation}
g_{\sigma}(s,\bm R;\omega_n,\omega_n')=
g_{\sigma}^{(0)} (\omega_n)\frac{1}{T}\delta_{n,n'}+g_{\sigma}^{(1)}(s,\bm R;\omega_{n},\omega_{n}-\omega_{m})\frac{1}{T}\delta_{n-n',m},
\end{equation}
where $g^{(0)}(\omega_n)$ is the unperturbed Green function and $g^{(1)}(s,\bm R;\omega_{n},\omega_{n}-\omega_{m})$ is the first order part in $\bm T_m$.
Using the above separation, the electric current is given by 
\begin{align}
\bm j (\bm R)&=
\sum_{\sigma=\pm}
\sum_{\omega_n} 
\pi N(\mu_\sigma)e T 
\int d s^2 
\bm v_{\sigma}(s)g_{\sigma}^{(1)}(s,\bm R;\omega_{n},\omega_{n}-\omega_{m}),
\end{align}
where $N(\epsilon)$ is the density of states and the sum over $\sigma$ indicates 
including the contributions of the majority and minority bands.
We assume that $\frac{ \epsilon_F\tau_{imp}}{\hbar} (k_F l_{sk})^{-2} \ll 1$,
which corresponds to the $\omega_c \tau_{imp}  \ll 1 $ where $\omega_c$ is the cyclotron frequency of the field 
$\bm \nabla \times \bm u (\bm R)$.
We then  calculate perturbatively and obtain  
 \begin{align}
\bm  j(\bm R)&=
-\frac{e^2}{3}\sum_{\sigma=\pm}N(\mu_\sigma) q_\sigma v_{\sigma}^2 
\bm {T_m}
 \frac{\omega_m}{\omega_m+1/\tau_{imp}}
+\frac{e^3}{3m}\sum_{\sigma=\pm}N(\mu_\sigma) v_{\sigma}^2 
\left(\bm{\nabla}\times\bm u (\bm R)\right)\times \bm T_m 
\frac{\omega_m}{(\omega_m +1/\tau_{imp})^2}
,\\
&\approx
-N(\epsilon_F)\frac{2e^2JM}{m}\bm {T_m}
 \frac{\omega_m}{\omega_m+1/\tau_{imp}}
+
N(\epsilon_F)
\frac{4e^3 \epsilon_F}{3m^2} 
\left(\bm{\nabla}\times\bm u (\bm R)\right)\times \bm T_m
\frac{\omega_m}{(\omega_m +1/\tau_{imp})^2} 
,
\end{align}
where we used that $JM/\epsilon_F \ll 1$. 

In the case of $\tau_{imp}/\tau_{sk} \ll1$, we take the dc-limit, perform analytical continuation, and then obtain  

 \begin{align}
\bm  j(\bm x, t)&=
N(\epsilon_F)
\left(\frac{2e^2JM}{m}\right)
\tau_\mathit{imp}
\left[ \mathbf{E}^{em}(x_0)
	   -\frac{m \alpha_{so}}{|e|}\frac{\partial}{\partial t}\bm{n}(x_0)
\right]
-
N(\epsilon_F)
\left(
\frac{4e^3 \epsilon_F}{3m^2} 
\right)
\tau_\mathit{imp}^2
\left[\mathbf{B}^{em}(x_0)
		+\frac{m \alpha_{so}}{|e|}\bm{\nabla}\times\bm{n}(x_0)\right]
\times
\left[ \mathbf{E}^{em}(x_0)
	   -\frac{m \alpha_{so}}{|e|}\frac{\partial}{\partial t}\bm{n}(x_0)
\right],\nonumber\\
&=
N(\epsilon_F)\left(\frac{2e^2JM}{m}\right)
\tau_\mathit{imp}
\bm{\mathcal{E}}^{\mathit{eff}}(x_0)
-N(\epsilon_F)\left(\frac{4e^3 \epsilon_F}{3m^2}\right) 
\tau_\mathit{imp}^2
\bm{\mathcal{B}}^{\mathit{eff}}(x_0)
\times
\bm{\mathcal{E}}^{\mathit{eff}}(x_0),
\end{align}
which is applicable at points $x = (\bm x, t)$ located within a range of $(\bm k_F ^{-1}$, $\hbar/\epsilon_F)$  from $x_0 = (\bm x_0, t_0)$. By evaluating at any selected point, we obtain Eq. (\ref{eq:current1}).   

\appendix
\section{The model of a skyrmions-merging process}
\label{model}
\begin{figure}[b]
	\begin{minipage}[b]{0.2\columnwidth}
		\begin{center}
			\includegraphics[ clip, width=0.8\columnwidth]{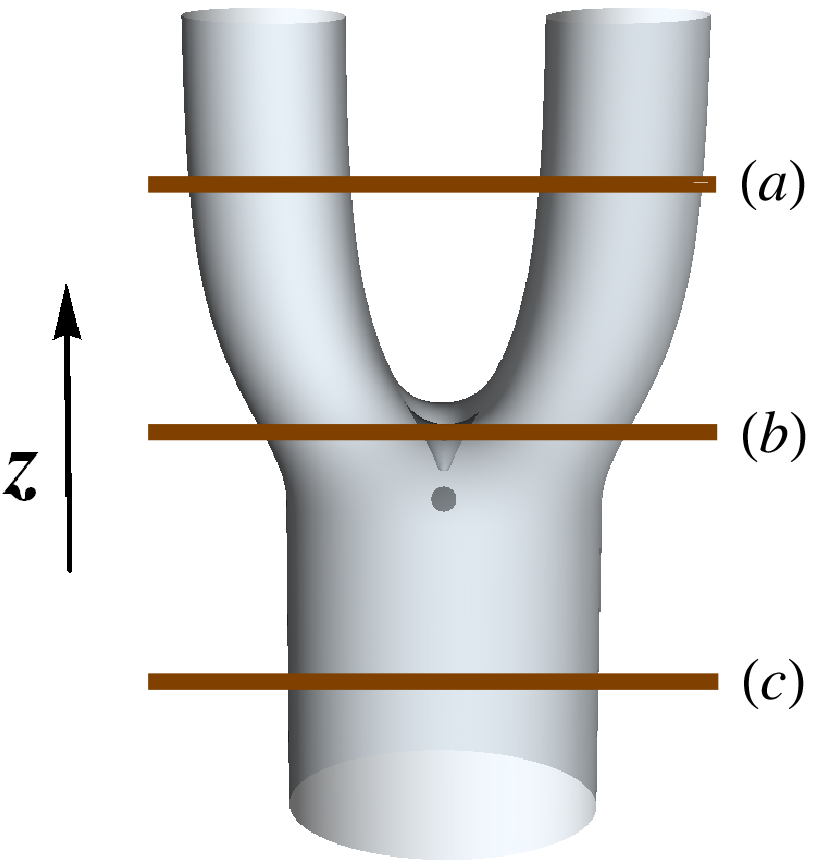}
		\end{center}
	\end{minipage}%
	\begin{minipage}[b]{0.8\columnwidth}
		\begin{center}
			\includegraphics[bb= 0 20 696 246 clip, width=\columnwidth]{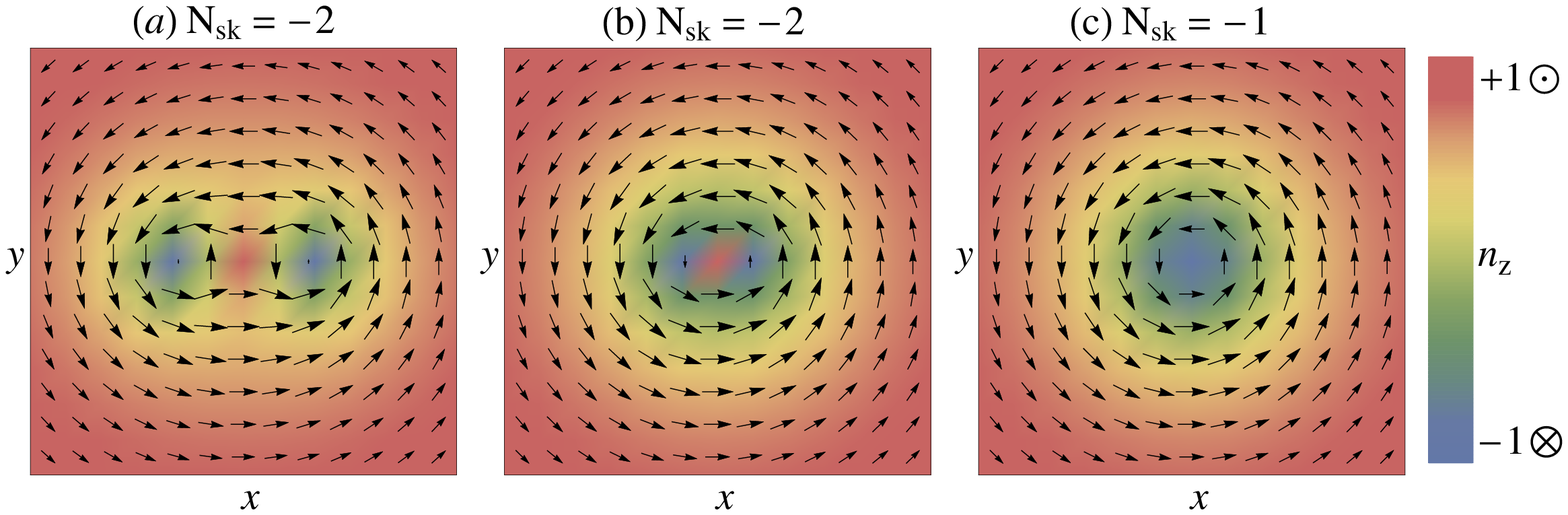}		
		\end{center}
	\end{minipage}\label{fig:append}
		\caption{(Color online) The spin configuration of the exchange field in our model at a certain time.  The arrows indicate the direction of the in-plane component of spins, and the color indicates the normal component to each plane shown in the left figure. The planes (a) and (b) locate above the merging point, and the skyrmion number (written  as $N_{sk}$) over each plane is $-2$. On the other hand, the plane (c) locates below the merging point and $N_{sk}=-1$ over the plane.  
}
\end{figure}

In Appendix\ref{model}, we show the detail of the model for the exchange field.
For the configuration of skyrmions, we have used a soliton solution of the $O(3)$ nonlinear sigma model in two-dimensional space\cite{Rajaraman1983} whose Lagrangian density is given by  
\begin{align}
\mathcal{L}=\frac{1}{2}{\partial _{\alpha}}\bm n(\bm r) \cdot{\partial_{\alpha}}\bm n(\bm r). \label{eq:sigma}
\end{align}
Here $|\bm n(\bm r)|^2=1$ and $\alpha ={x, y}$. 
We can rewrite $\bm n(\bm r)$ by using a complex scalar field $u(\bm r)$ as
\begin{align}
\bm n(\bm r) =\frac{1}{1+|u|^2}(1\ \ u^*)
\bm \sigma \begin{pmatrix}1\\ u\end{pmatrix}. 
\end{align} 
Here we introduce a topological number for a $xy$ plane as    
\begin{align}
N_{sk}=\int d^{2}x \frac{1}{4\pi}\bm n\cdot (\partial_{x}\bm n\times \partial_{y}\bm n).
\end{align}
We can define a topological number obtained by integration over $S^{2}$ as well. 
The configuration of a skyrmion is a saddle-point solution in Eq. (\ref{eq:sigma}), and
one skyrmion state with skyrmion number $-1$ can be decribed as
\begin{align}
 u=\frac{R e^{i \gamma}}{(x-iy)-(x_c -iy_c)},
\end{align}
where $(x_c, y_c)$ is the center of the skyrmion, $R$ is the length scale of the skyrmion radius, $\gamma$ is the skyrmion helicity, and the boundary condition is given as $\bm n (|\bm r|\rightarrow\infty)=(0,0,1)^\mathrm{T}$. 
In our three-dimensional model, we consider the following field written in terms of $u$ as 
\begin{align}
 u=\frac{Re^{i \gamma}}{(x-iy)-w(z-v_mt)}+\frac{Re^{i \gamma}}{(x-iy)+w(z-v_mt)},
\end{align}
where $v_m$ is the velocity of a monopole and $2w(z-v_mt)$ is the distance between the two skyrmions.
Here we assumed that $w(Z)\propto \tanh Z $ for $Z>0$, and $w(Z)=0 $ for $Z=0$ for simplicity.
The skyrmion number evaluated in a $xy$ plane above the monopole is $-2$, while the number evaluated in a $xy$ plane below the monopole is $-1$, as indicated in Fig.\ref{fig:append} $\cdot\  1$.
Except for the change in the distance around the merging point, we assume the homogeneous structure in the $z$ direction. 
There is a point defect at the point $x=y=0$ and $z-v_m t=0$ ,where $n_z$ is discontinuous, and the surrounding configuration is topologically non-trivial; it is a hedgehog monopole. 

\end{document}